\documentclass[%
 reprint,
 amsmath,amssymb,
 aps,
]{revtex4-2}

\usepackage{graphicx}% Include figure files
\usepackage{dcolumn}% Align table columns on decimal point
\usepackage{bm}% bold math
\usepackage{xcolor}
\usepackage[colorlinks, citecolor = blue, urlcolor = blue, linkcolor = blue]{hyperref}% add hypertext capabilities

\usepackage{soul}  % For strikethrough
\usepackage{ulem}

\begin{document}

\preprint{APS/123-QED}

\title{Formation of Complex Discrete Time Crystals with Ultracold Atoms}% Force line breaks with \\

\author{Weronika Golletz$^{1,2,}$}\thanks{weronika.golletz@protonmail.com}
\author{Krzysztof Sacha$^{1,}$}\thanks{krzysztof.sacha@uj.edu.pl}
\affiliation{%
 $^1$Faculty of Physics, Astronomy and Applied Computer Science, Institute of Theoretical Physics, Jagiellonian University, Lojasiewicza 11, PL-30-348 Krakow, Poland \\
 $^2$Doctoral School of Exact and Natural Sciences, Jagiellonian University, Lojasiewicza 11, PL-30-348 Krakow, Poland
}%

\date{\today}

\begin{abstract}
We study discrete time crystal formation in a system driven periodically
by an oscillating atomic mirror, consisting of two distinct ultracold atomic clouds in the presence of a gravitational field. 
The intra-species interactions are weak and attractive, while the inter-species interactions are infinitely strong and repulsive. 
The clouds are arranged in a one-dimensional stack, where the bottom cloud bounces on an oscillating atomic mirror, which effectively acts as a driving force for the upper cloud due to the infinite inter-species repulsion. 
Using a Jastrow-like variational ansatz for the many-body wavefunction, we show that sufficiently strong attractive intra-species interactions drive each subsystem to spontaneously break discrete time translation symmetry, resulting in the formation of a \textit{complex discrete time crystal} evolving with a period different than the driving period.
Since the bottom cloud serves as the effective periodic driving for the upper cloud, this leads to a cascade of spontaneous symmetry breaking. With increasing intra-species interactions, we first observe a pronounced effect of spontaneous time translation symmetry breaking in the upper cloud, followed by a similar effect in the lower atomic cloud.
\end{abstract}

\maketitle

%\tableofcontents

\section{\label{sec:introduction}Introduction}
In 2012, Frank Wilczek proposed that a quantum many-body system in its lowest energy state could spontaneously break continuous time translation symmetry, forming a time crystal, analogous to the formation of spatial crystals~\cite{Wilczek2012}. 
However, it was later shown that such time crystals cannot exist in equilibrium systems with two-body interactions~\cite{Bruno2013b, Watanabe2015}. 
Instead, discrete time crystals (DTCs) can emerge in periodically driven many-body systems, where the system exhibits a response at an integer multiple of the driving period, spontaneously breaking discrete time translation symmetry~\cite{Sacha2015,Khemani16,ElseFTC,Yao2017,Zhang2017,Choi2017,Sacha2017rev,Pal2018,Rovny2018,Autti2018,
Smits2018,Kessler2020,SachaTC2020,Taheri2020,Autti2021,GuoBook2021,Kyprianidis2021,Xu2021,Randall2021,Autti2022,Mi2022,Frey2022,Zaletel2023,AEK2024,Bao2024,Kazuya2024,Liu2024,Liu2024a,Paulino2025,Makinen2025}.

In this paper, we focus on DTCs which can be realized in an ultracold atomic cloud~\cite{Sacha2015}. 
In such a setup, attractively interacting atoms bounce resonantly on an oscillating mirror
in the presence of the gravitational field.
It has been shown that this system can host big DTCs, where spontaneous discrete time translation symmetry breaking leads to periodic evolution with a period much longer than the driving period,
reaching up to $s=100$ times the driving period~\cite{Giergiel2018a,Giergiel2020}. 

\begin{figure}[t]
\includegraphics[]{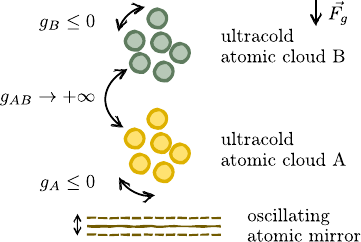}% Here is how to import EPS art
\caption{Sketch of the system under study. It consists of two distinct ultracold atomic clouds, 
$A$ and $B$, arranged in a one-dimensional stack. The system is periodically driven by an oscillating atomic mirror,
with period $T=2\pi/\omega$, in the presence of a gravitational field~$\vec{F}_g$. 
Intra-species interactions, 
$g_{\sigma}$ for $\sigma \in \{A,B\}$, are weakly attractive, while inter-species interactions are infinitely repulsive, $g_{AB} \rightarrow \infty$. This strong repulsion prevents overlap between clouds and causes cloud $A$ to effectively act as a driving force for cloud $B$.}
\label{fig:setup}
\end{figure}

Here, we take a step in a different direction -- instead of focusing on big DTCs, we explore the impact of spontaneous breaking of discrete time translation symmetry in a more complex system. 
Rather than a single bouncing ultracold atomic cloud, we consider a two-component system consisting of two interacting atomic clouds, see Fig.~\ref{fig:setup}. 
Atoms within each cloud interact attractively, while atoms from different clouds experience an infinitely strong inter-species repulsion, preventing spatial overlap. This strong repulsion effectively couples their motion, as the bottom cloud bounces on the oscillating mirror and indirectly drives the upper cloud. 
A key question we address is how discrete time translation symmetry breaking affects the periodicity of this coupled system. 
For a single ultracold cloud, it is known that increasing the attractive interaction strength induces DTC formation~\cite{Sacha2015}.
Our goal is to analyze whether the presence of the second cloud suppresses symmetry breaking or instead leads to the emergence of a complex DTC.

The paper is organized as follows.
In Sec.~\ref{sec:system}, we introduce the system, discuss the formation of DTCs in the case of two independent ultracold atomic clouds (Sec.~\ref{sec:nonint-clouds}), and provide the theoretical framework needed to analyze the behavior of two interacting ultracold clouds (Sec.~\ref{sec:int-clouds}).
In Sec.~\ref{sec:meth-results}, we present a detailed analysis of spontaneous discrete time translation symmetry breaking in the system and discuss our results.
Finally, we summarize our work in Sec.~\ref{sec:summary}.

%%%%%%%%%%%%%%%%%%%%%%%%%%%%%%%%%%%
%%%%%%%%%%%%%%%%%%%%%%%%%%%%%%%%%%%
\section{\label{sec:system}Model}
We consider a system of two distinct ultracold bosonic atomic clouds,
$A$ and $B$, arranged in a one-dimensional (1D) vertical stack.
The system is periodically driven by an oscillating atomic mirror with period $T=2\pi/\omega$, 
in the presence of a gravitational field $\vec{F}_g$, 
as illustrated in Fig.~\ref{fig:setup}.
We assume that all particles have identical masses, which may correspond to atoms of the same species but in different internal states for each cloud~\cite{Sowiński2019}.
Atoms within each cloud interact weakly and attractively, characterized by an intra-species interaction strength $g_{\sigma} \leq 0$ , where $\sigma \in \{A,B\}$.
In contrast, inter-species interactions are infinitely repulsive, $g_{AB} \rightarrow \infty$.
This strong repulsion causes cloud $A$ to effectively act as a driving force for cloud $B$.

Since the system is periodically driven, 
we look for a kind of stationary states that evolve periodically in time.
These states correspond to eigenstates of the Floquet Hamiltonian,
expressed in the oscillating mirror's reference frame~\cite{footnote3}:
\begin{eqnarray}
    \hat{\mathcal{H}} = \hat{H}_A + \hat{H}_B + \hat{H}_{AB} - i \partial_t
    \label{eq:hamiltonian-full}
\end{eqnarray}
where $\hat{H}_{\sigma}$ is the Hamiltonian of each individual atomic cloud $\sigma$,
and $\hat{H}_{AB}$ describes the interaction between the two clouds:
\begin{align}
    \hat{H}_{\sigma} =&\sum^{N_{\sigma}}_{j=1} \left(\frac{-\partial^2_{x_j}}{2} + x_j + \lambda x_j \cos(\omega t) \right) \\&+ \frac{g_{\sigma}}{2}\sum^{N_{\sigma}}_{j\ne k=1}\delta(x_j-x_k),\\
    \hat{H}_{AB} =& \frac{g_{AB}}{2}\sum^{N_A}_{j=1}\sum^{N_B}_{k=1}\delta(x_j-y_k).
\end{align}
The mirror oscillates with frequency $\omega$ and amplitude determined by $\lambda$~\cite{footnote3},
while interactions between atoms are modeled by a Dirac-delta contact potential $\delta(x)$.
Each cloud consists of $N_{\sigma}$ particles.
Throughout this paper, we employ gravitational units where length, energy, and time are given by $l_0 = (\hbar^2/m^2 g)^{1/3}$, $E_0 = mg l_0$ and $t_0 = \hbar/mgl_0$, respectively.
Here, $g$ is the gravitational acceleration and $m$ the atomic mass.
A classical analysis of this system was previously given in Ref.~\cite{Golletz2022}.

%%%%%%%%%%%%%%%%%%%%%%%%%%%%%%%%%%%%%%%
%%%%%%%%%%%%%%%%%%%%%%%%%%%%%%%%%%%%%%%
%%%%%%%%%%%%%%%%%%%%%%%%%%%%%%%%%%%%%%%
\subsection{Two independent atomic clouds}\label{sec:nonint-clouds}
In the absence of inter-species interactions, $g_{AB}=0$, 
the system described by the Hamiltonian~\eqref{eq:hamiltonian-full} decouples into two independent subsystems,
each evolving under its own many-body Floquet Hamiltonian given by $\hat{H}_{\sigma} \rightarrow \hat{\mathcal{H}}_{\sigma} = \hat{H}_{\sigma} - i \partial_t$.
We focus on a regime where each subsystem is resonantly driven, 
meaning that the driving frequency $\omega$ is an integer multiple of the frequency $\Omega_{\sigma}$ of an unperturbed motion of a particle, $s_{\sigma} \Omega_{\sigma} = \omega$ where $s_\sigma$ is an integer.
Under these conditions and for sufficiently small $\lambda$, the system follows a resonant trajectory, avoiding chaotic behavior and maintaining stable, controlled dynamics.
Such a setup corresponds to an $s_{\sigma} : 1$ resonance between the external driving and the particle motion~\cite{Buchleitner1995,Lichtenberg1992}.
Under this condition, each subsystem can be effectively described by a Bose-Hubbard Hamiltonian~\cite{Sacha2015,Sacha15a,SachaTC2020,Wang2020,Kuros2020,Wang2021}: 
\begin{equation}
    \hat{\mathcal{H}}_{\sigma}^{\rm eff} = -\frac{J^{\sigma}}{2} \sum_{\langle j,k \rangle} \left(\hat{a}^{\sigma\dagger}_j \hat{a}_k^{\sigma}  + \hat{a}^{\sigma\dagger}_k \hat{a}_j^{\sigma}\right) + \frac{1}{2} \sum_{j,k} U_{jk}^{\sigma} \hat{a}^{\sigma\dagger}_k \hat{a}_j^{\sigma\dagger} \hat{a}^{\sigma}_j \hat{a}_k^{\sigma},
    \label{eq:Bose-Hubbard}
\end{equation}
where $\hat{a}^{\sigma}$ are bosonic annihilation operators,
the tunneling amplitude is given by
$J^{\sigma} = 
-\frac{2}{s_{\sigma} T} \int_0^{s_{\sigma} T} dt \int_0^{\infty} dx \, 
w_j^{\sigma\,*}(x,t)(\hat{H}_{\sigma}-i\partial_t) w_j^{\sigma}(x,t)$
and interaction coefficients are determined by $U_{jk}^{\sigma} = \frac{2 g_{\sigma}}{s_{\sigma} T} \int_0^{s_{\sigma} T} dt \int_0^{\infty} dx \, |w_j^{\sigma}(x,t)|^2 |w_k^{\sigma}(x,t)|^2$ for $j\neq k$.
The on-site interaction coefficients $U_{jj}^{\sigma}$ follow a similar expression but are smaller by a factor of two.
The Bose-Hubbard Hamiltonians, $\hat{\mathcal{H}}_{\sigma}^{\rm eff}$, are
written in terms of a time-periodic basis of single-particle wavefunctions $w_j^{\sigma}(x,t)$.
These wavefunctions, known as Wannier functions -- analogous to those in condensed matter physics~\cite{Ashcroft1976} -- form localized wavepackets that propagate along classical resonant trajectory with a period $T_{\sigma}=s_{\sigma} 2\pi / \omega$.
For a given resonance, the set of Wannier functions spans the resonant Hilbert subspace, i.e., the single-particle wavefunction can be expressed as:
\begin{equation}
    \psi_{\sigma}(x,t) = \sum_{j=1}^{s_{\sigma}} c_j^{\sigma} w_j^{\sigma}(x,t).
    \label{eq:single-particle-state}
\end{equation}

In the regime of weak intra-species interactions, $g_{\sigma} \rightarrow 0$, the lowest quasi-energy state of Hamiltonian $\hat{\mathcal{H}}_{\sigma}^{\rm eff}$ corresponds to a Bose-Einstein condensate (BEC),
\begin{equation}
    \Psi^{\rm GS}_{\sigma}(\vec{x},t) = \Pi^{N_{\sigma}}_{j=1} \psi_{\sigma}(x_j,t),
    \label{eq:GS-weak}
\end{equation} 
where $\vec{x} = (x_1,\hdots,x_{N_{\sigma}})$, 
and all atoms occupy a state~\eqref{eq:single-particle-state} 
with coefficients $\forall_j\,c_j^{\sigma} = 1/\sqrt{s_{\sigma}}$ which is a uniform superposition of the Wannier functions.
The state $\Psi^{\rm GS}_{\sigma}(\vec{x},t)$ evolves with the driving period $T=2\pi/\omega$. 
Although each Wannier state is a periodic function with period $T_\sigma=s_\sigma T$, 
their uniform superposition remains periodic with period $T$ due to the relation $w^\sigma_j(x,t+T)=w^\sigma_{j+1}(x,t)$.

However, when the intra-species interactions become sufficiently strong and attractive $|g_{\sigma}|>0$, a mean-field approach shows that it becomes energetically favorable for all atoms to localize in a single Wannier state $w_j^{\sigma}(x,t)$~\cite{Wuster2012,Wuster2024,Sacha2015}, i.e., only one coefficient in~\eqref{eq:single-particle-state} is nonzero.
On the other hand, the Floquet state must preserve the discrete time translation symmetry of the Hamiltonian \eqref{eq:hamiltonian-full}.
As a result, the lowest quasi-energy state takes the form of a macroscopic superposition of BECs~\cite{Zin2008, Oles2010}: 
\begin{equation}\Psi^{\rm GS}_{\sigma}(\vec{x},t) \propto \sum_{k=1}^{s_{\sigma}}\Pi^{N_{\sigma}}_{j=1} w_k^{\sigma}(x_j,t).
\label{eq:GS-strong-symmatric}
\end{equation} 
This superposition state~\eqref{eq:GS-strong-symmatric} is extremely sensitive to perturbations. 
For instance, measuring the position of a single atom induces a collapse into one of the product states:
\begin{equation}
    \Psi^{\rm GS}_{\sigma}(\vec{x},t) \rightarrow \Psi_{\sigma}(\vec{x},t) \approx \Pi^{N_{\sigma}}_{j=1} w_k^{\sigma}(x_j,t).
    \label{eq:GS-strong-notsymmatric}
\end{equation} 
The specific Wannier state $w_k^{\sigma}(x,t)$ into which the system collapses depends on the measurement outcome. 
The resulting state $\Psi_{\sigma}(\vec{x},t)$~\eqref{eq:GS-strong-notsymmatric} evolves with a period $s_{\sigma}$ times longer than that of $\Psi^{\rm GS}_{\sigma}(\vec{x},t)$~\eqref{eq:GS-strong-symmatric}.
This corresponds to spontaneous discrete time translation symmetry breaking, leading to the formation of a DTC, where the system exhibits periodicity that is an integer multiple of the driving period~\cite{Sacha2015, Khemani16, Else16}.

Since in the regime $g_{AB}=0$, the two atomic clouds are independent, each undergoes its own DTC phase transition for sufficiently strong intra-species interactions $|g_{\sigma}|>0$. 
In this work, we explore cases where $s_{A}=2$, corresponding to a period-doubling DTC, and 
$s_B=3$, leading to a period-tripling DTC.

%%%%%%%%%%%%%%%%%%%%%%%%%%%%%%%%%%%%%%%
%%%%%%%%%%%%%%%%%%%%%%%%%%%%%%%%%%%%%%%
%%%%%%%%%%%%%%%%%%%%%%%%%%%%%%%%%%%%%%%
\subsection{Two interacting atomic clouds}\label{sec:int-clouds}
In the limit of infinitely repulsive inter-species interactions, $g_{AB}\rightarrow \infty$,
non-trivial correlations arise between particles from different subsystems,
making the system non-separable.
The bottom cloud $A$ effectively acts as a driving force for cloud $B$.
To simplify the analysis,
we eliminate the infinite interaction term~$\hat{H}_{AB}$ from the 
Hamiltonian~\eqref{eq:hamiltonian-full}
by incorporating the inter-species interactions into an antisymmetric two-particle basis:
\begin{align}
    &\Phi(x,y;t) = \mathcal{A}[\psi_A(x,t)\psi_B(y,t)] \label{eq:antisymmetric-basis}\\
                &= \mathcal{N} \sum_{j=1}^{s_A} \sum_{k=1}^{s_B} c_j^A c_k^B (w_j^A(x,t)w_k^B(y,t) - w_j^A(y,t)w_k^B(x,t)),   \notag  
\end{align}
where $\psi_{\sigma}(x,t)$ is defined in~\eqref{eq:single-particle-state}, $\mathcal{N}$ is the normalization constant, and
$\mathcal{A}[\cdot]$ is the antisymmetrization operator.
We then use this basis to construct the many-body Floquet state $\Psi(\vec{x},\vec{y};t)$:
\begin{equation}
    \Psi(\vec{x},\vec{y};t) = \Pi^{N_A}_{m=1} \Pi^{N_B}_{m'=1} \Phi(x_m,y_{m'};t),
    \label{eq:Jastrow}
\end{equation}
which takes the form of a Jastrow-like ansatz~\cite{Jastrow1955} 
with variational parameters 
$\{\vec{c}_A,\vec{c}_B\} = \{c_1^A,\hdots,c_{s_A}^A;c_1^B,\hdots,c_{s_B}^B\}$.
The ansatz is symmetric under the exchange of any two particles of the same type, i.e., $x_j \leftrightarrow x_k$ and $y_j \leftrightarrow y_k$,
as required for bosonic atoms. 
Moreover, the wavefunction~\eqref{eq:Jastrow} vanishes when two atoms from different subsystems coincide, i.e., $\Psi(\vec{x},\vec{y};t)\rvert_{x_j = y_k}=0$,
which incorporates the effects of infinitely repulsive inter-species interactions.
As a result, we effectively remove $\hat{H}_{AB}$ from the Hamiltonian~\eqref{eq:hamiltonian-full}, yielding:
\begin{eqnarray}
    \hat{\mathcal{H}} = \hat{H}_A + \hat{H}_B - i \partial_t.
    \label{eq:hamiltonian-without-int}
\end{eqnarray}

To find the lowest quasi-energy state of~\eqref{eq:hamiltonian-without-int} within the resonant Hilbert subspace, we minimize the expectation value of \eqref{eq:hamiltonian-without-int} 
using the variational state~\eqref{eq:Jastrow},
optimizing the coefficients $\{\vec{c}_A,\vec{c}_B\}$.
For a given intra-species interaction strength $g_{\sigma}$, 
the quasi-energy functional is defined as
\begin{equation}
    E[\Psi(\vec{c}_A,\vec{c}_B)] = \frac{\int_0^{T_{\rm tot}} dt \int_0^{\infty}d\vec{x}\,d\vec{y}\, \Psi^*(\vec{x},\vec{y};t)\hat{\mathcal{H}}\Psi(\vec{x},\vec{y};t)}{\int_0^{T_{\rm tot}} dt \int_0^{\infty}d\vec{x}\,d\vec{y}\, \Psi^*(\vec{x},\vec{y};t)\Psi(\vec{x},\vec{y};t)}
    \label{eq:quasienergy-functional}
\end{equation}
where $T_{\rm tot} = s_A s_B T$ is the total period of the system's motion, i.e., the longest period attainable in the resonant Hilbert subspace spanned by the Wannier states.
% Since the Jastrow ansatz~\eqref{eq:Jastrow} has periodicity $T_{\rm tot}=s_As_B T$,
% we take a time average of the functional over this period.

%%%%%%%%%%%%%%%%%%%%%%%%%%%%%%%%%%%%%%%
%%%%%%%%%%%%%%%%%%%%%%%%%%%%%%%%%%%%%%%
%%%%%%%%%%%%%%%%%%%%%%%%%%%%%%%%%%%%%%%

\section{Methodology and Results}\label{sec:meth-results}
% Before minimization of quasi-energy functional~\eqref{eq:quasienergy-functional},
% we need to choose the proper parameter of our system, such as $\lambda$ and amplitude $\omega$ of the oscillating mirror.
% We select the system parameters based on classical phase space portraits for single-particle system and a given resonance $s_{\sigma}$. Here, we summarize the most important contstraint of those parameters, and for deeper understanding we recommend Refs.~\cite{Sacha2017rev,Giergiel2018,Giergiel2018a,Giergiel2020,SachaTC2020}. 

Minimizing the quasi-energy functional~\eqref{eq:quasienergy-functional} becomes computationally challenging because its complexity grows exponentially with the total number of particles, 
$N=\sum_{\sigma}N_\sigma$.
To keep the problem manageable, 
we study a system with two ultracold clouds, each containing $N_A=N_B=2$ particles, in resonances $s_A=2$ and $s_B=3$.
In addition, we use a Gaussian approximation for the Wannier states $\{w_j^{\sigma}(x,t)\}$,
by replacing each $w_j^{\sigma}(x,t)$ with
\begin{align}
    g_j^{\sigma}(x;\mu_j^{\sigma}(t),\nu_j^{\sigma}(t),&k_j^{\sigma}(t)) = \left(\frac{1}{2\pi (\nu^{\sigma}_j)^2} \right)^{1/4} \label{eq:gaussians}\\ 
    &\times \exp\left(-\frac{(z-\mu_j^{\sigma}(t))^2}{4(\nu_j^{\sigma}(t))^2} + i k_j^{\sigma}(t) x \right), \notag 
\end{align}
where the Gaussian function is characterized by its mean $\mu_j^{\sigma}(t)$,
standard deviation $\mu_j^{\sigma}(t)$ and momentum $k_j^{\sigma}(t)$ at a given time $t$.
The Gaussian approximation is crucial in our calculations because it allows us to perform all integrals in (\ref{eq:quasienergy-functional}) analytically. We find the optimal parameters $\chi_j^{\sigma}(t) = \{\mu_j^{\sigma}(t),\nu_j^{\sigma}(t),k_j^{\sigma}(t)\}$ by maximizing
\begin{equation}
    F_j^{\sigma}(t) = \underset{\mu,\nu,k}{\rm max} \left(|\langle g_j^{\sigma}(x;\chi_j^{\sigma}(t)) | w_j^{\sigma}(x,t) \rangle|^2\right)
\end{equation}
for each time $t$.
The average squared overlap over a total period
$\bar{F}_j^{\sigma} = \frac{1}{T_{\rm tot}} \int_0^{T_{\rm tot}} F_j^{\sigma}(t)\, dt$,
quantifies the effectiveness of the Gaussian approximation. For the parameters chosen in our paper, the average squared overlaps are $\bar{F}_j^A=0.71$ and $\bar{F}_j^B=0.70$.

\begin{figure}[t]
\includegraphics[]{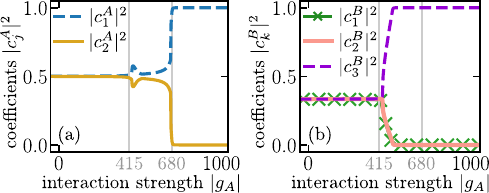}
\caption{The lowest quasi-energy solutions of Floquet Hamiltonian~\eqref{eq:hamiltonian-without-int}, within the resonant Hilbert subspace, take the form of  Jastrow-like ansatz~\eqref{eq:Jastrow} with optimized coefficients 
$\{c^A_1,c_2^A;c_1^B,c_2^B,c_3^B\}$, obtained by minimizing the quasi-energy functional~\eqref{eq:quasienergy-functional}. 
For clarity, the coefficients corresponding to the resonance $s_A=2$ (dashed blue and solid yellow lines)
are shown in Panel~(a), while those for the resonance $s_B=3$ (dashed violet, solid pink, and solid green with crosses) are displayed in Panel~(b).
For $|g_A| \lesssim 415$, the coefficients within each resonance remain equal, i.e., $|c_j^A|^2=1/2$ for $j=1,2$ and $|c_k^B|^2=1/3$ for $k=1,2,3$, indicating that the system respects the discrete time translation symmetry. 
In this regime, the many-body wavefunction~\eqref{eq:Jastrow} evolves with the same period as the driving period $T=2\pi/\omega$.
For $|g_A| \gtrsim 415$, the coefficients within each resonance become unequal, signaling that the many-body wavefunction breaks the discrete time translation symmetry.
As a results, the system starts moving with a period $T_{\rm tot} = s_A s_BT$,  which is longer than the driving period $T$.
The Hamiltonian~\eqref{eq:hamiltonian-without-int} parameters are: 
$\omega = 1.4$, $\lambda = 0.07$, $s_A=2$, $s_B=3$ and $N_A=N_B=2$.
}
\label{fig:coefficients}
\end{figure}

Next, we minimize the quasi-energy functional~\eqref{eq:quasienergy-functional}.
This minimization is performed under the normalization constraints:
$\langle \Psi | \Psi \rangle = \frac{1}{T_{\rm tot}} \int_0^{T_{\rm tot}} dt \int_0^{\infty} d\vec{x}\, d\vec{y}\, |\Psi(\vec{x},\vec{y};t)|^2 = 1$ and
$\sum_{j=1}^{s_{\sigma}}|c_j^{\sigma}|^2=1$.
In Fig.~\ref{fig:coefficients}, we show how the optimized coefficients  $\{\vec{c}_A,\vec{c}_B\}$ change for different intra-species interaction strength $g_A$, while keeping $g_B=0$.
To clearly distinguish the behavior in each subsystem, the optimized coefficients are plotted separately for each resonance $s_{\sigma}$.
For $|g_A| \lesssim 415$, the functional~\eqref{eq:quasienergy-functional} is minimized by a uniform solution, with $|c_j^A|^2 = 1/2$ for $j=1,2$ and $|c_k^B|^2 = 1/3$ for $k=1,2,3$. 
In this regime, the many-body state $\Psi(\vec{x},\vec{y};t)$~\eqref{eq:Jastrow} evolves with the same period as the oscillating mirror, $T=2\pi/\omega$, preserving discrete time translation symmetry.
For $|g_A| \gtrsim 415$, the uniform solution no longer minimizes the energy functional, giving rise to symmetry-broken states. These states evolve with a period $T_{\rm tot} = s_As_BT$,
indicating the simultaneous breaking of discrete time translation symmetry in both subsystems and the emergence of a complex DTC.
Within the range $|g_A| \in [415,680]$, the lowest quasi-energy state exhibits a non-uniform distribution of the coefficients $\{\vec{c}_A,\vec{c}_B\}$ but the breaking of discrete time translation symmetry is not dramatically evident in the lower atomic cloud as $|c^A_1|^2\approx|c^A_2|^2$. This implies that the lower cloud, which drives the upper atomic cloud, nearly follows the mirror’s oscillations, whereas in the upper driven cloud the symmetry breaking is evident.
For $|g_A| \gtrsim 680$, the symmetry-broken state is characterized by a single Wannier state $w_j^{\sigma}(x,t)$ in each subsystem, meaning that only one coefficients $\vec{c}_{\sigma}$ remains non-zero for each $\sigma \in \{A,B\}$.

It is worth noting that a total attractive intra-species interaction strength satisfying $|g_A+g_B| = {\rm constant} > 415$ is sufficient to observe the emergence of the complex DTC.
In Figure~\ref{fig:coefficients}, we present results where $|g_A|$ varies while keeping $g_B=0$.
However, the optimized coefficients $\{\vec{c}_A,\vec{c}_B\}$ are independent of whether we vary $|g_B|$ while keeping $g_A=0$, or simultaneously change both $g_A<0$ and $g_B<0$.
This behavior arises from the ansatz~\eqref{eq:Jastrow},
which enforces symmetrization under the exchange of any two particles of the same type $\sigma$.
As long as the particle numbers remain equal, $N_A=N_B$,
the intra-species interaction contributions are identical. 
The key condition for spontaneous discrete time translation symmetry breaking is that the attractive intra-species interaction terms must dominate over the kinetic energy terms in the quasi-energy functional~\eqref{eq:quasienergy-functional}.

%\textcolor{red}{throught this paper we will recognize diffrent particle x y}

%\textcolor{red}{komentarz dlaczego gB0; czy jest wszystko jasne; dlaczego te wspolczynniki mowia wszystko; moze bardziej clear the state of the subsystem is characterize by the coeffitients, based on their values we can analyse the discrete time symmetry}

One might question why the intra-species interaction strength required to observe discrete time translation symmetry breaking, $|g_A|\sim 415$, is significantly larger compared to the interaction strength needed for discrete time translation symmetry breaking in the case of a single ultracold cloud bouncing on an oscillating mirror, as studied in Ref.~\cite{Sacha2015}.
This difference arises due to the use of the two-particle basis~\eqref{eq:antisymmetric-basis}, which incorporates infinite inter-species interactions through antisymmetrization. While this ensures that wavefunction~\eqref{eq:Jastrow} vanishes when two atoms from different subsystems coincide $\Psi(\vec{x},\vec{y};t)\rvert_{x_j = y_k}=0$, it also adds complexity by encompassing two independent particle configurations, where $x>y$ and $x<y$. 
As a consequence, the ansatz~\eqref{eq:Jastrow} includes all possible particle configuration sectors, such as those where $\forall_{m,m'} \, (x_m > y_{m'})$ or $\forall_{m,m'}\, (x_m < y_{m'})$, 
as well as sectors where particles of different species alternate e.g. $x_1 > y_1 > y_2 > x_1$.
Although these configuration sectors are not coupled by the Floquet Hamiltonian~\eqref{eq:hamiltonian-without-int}, the quasi-energy functional~\eqref{eq:quasienergy-functional} is minimized by averaging over all such sectors. 
Since the contribution of intra-species interaction terms varies across different configuration sectors, the effective interaction strength required for discrete time translation symmetry breaking is larger than it would be if the quasi-energy functional were minimized in a single sector, e.g. $\forall_{m,m'} \, (x_m < y_{m'})$.
Since our method accounts for all possible configuration sectors, the system’s behavior is fully encoded in the wavefunction with optimized coefficients. The relevant physical properties are then extracted by simulating measurements using the Monte Carlo method~\cite{Metropolis1953}, as described in the final paragraph of this section.
An alternative approach to this problem is to perform calculations within a single configuration sector by introducing Heaviside theta functions to avoid the need to account for all possible particle arrangements. However, this method is difficult to handle numerically, as it requires integrating Heaviside theta functions appearing in the quasi-energy functional.

\begin{figure}[!t]
\includegraphics[]{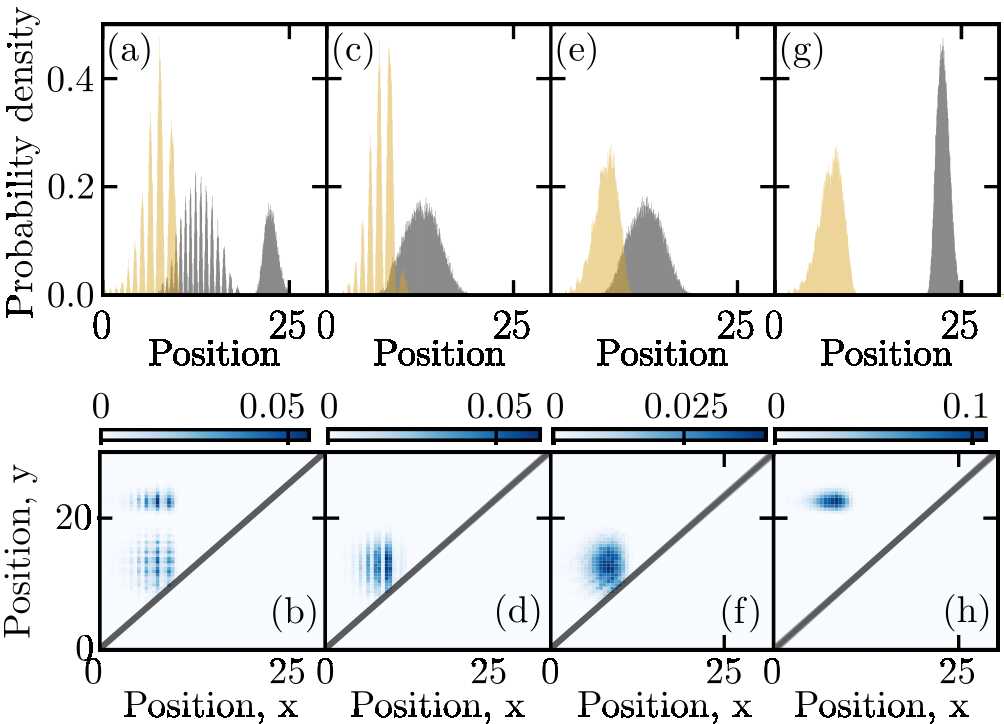}% Here is how to import EPS art
\caption{Probability densities extracted from the many-body wavefunction~\eqref{eq:Jastrow} under the constraint that all particles of type $A$ are positioned below those of type $B$.
The top panels show the marginalized probability distributions for $\{x\}$ (yellow bars) and $\{y\}$ (gray bars),  while the bottom panels display the corresponding joint distributions $(x_m,y_{m'})$ as 2D histograms. 
The data is obtained via Monte Carlo sampling with $50,000$ iterations, keeping only configurations $\{x_1,x_2,y_1,y_2\}$ satisfying $x_1,x_2 < y_1, y_2$.
Each column corresponds to a different many-body wavefunction optimized for intra-species interaction strength $g_A$, with values selected based on Fig.~\ref{fig:coefficients}. 
(a)-(b) Symmetry-preserving case $g_A=0$, where $\{|\vec{c}_A|^2,|\vec{c}_B|^2\} =\{1/2,1/2;1/3,1/3,1/3\}$ at time $t=0$. 
(c)-(d) Symmetry-broken regime for $|g_A| = 610$
with $\{|\vec{c}_A|^2,|\vec{c}_B|^2\} =\{0.46,0.54;1,0,0\}$ at time $t=0$.
(e)-(h) Another symmetry-broken case $|g_A| = 680$ with $\{|\vec{c}_A|^2,|\vec{c}_B|^2\} =\{1,0;1,0,0\}$, at time $t=0$ (e)-(f) and at time $t=2T$ (g)-(h).
}
\label{fig:probability}
\end{figure}

While the wavefunction that describes sectors with all possible particle configurations
is mathematically well-defined, 
it is difficult to prepare experimentally.
In typical setups, one species of particles is positioned above the other, as illustrated in Fig.~\ref{fig:setup}. To simulate such an experiment, we sample the many-particle probability density 
$|\Psi(\vec{x},\vec{y};t)|^2$ \eqref{eq:Jastrow} over different configurations of particles $\{\vec{x}, \vec{y}\}$ and extract only those configurations that satisfy the condition $\forall_{m,m'} \, (x_m < y_{m'})$,
ensuring that all particles of type $A$ are positioned below those of type $B$.
To achieve this, we perform a series of particle measurements using a Monte Carlo algorithm~\cite{Metropolis1953} based on a Markovian random walk that explores the configuration space~\cite{Press2007,Syrwid2016}.
We run $50,000$ iterations and apply a conditional selection criterion,
keeping only configurations where $x_1,x_2 < y_1, y_2$.
From this filtered subset, we construct the marginalized probability distribution for $x$ and $y$ by collecting all $x$-coordinates and $y$-coordinates separately.
The corresponding histograms are presented in Figure~\ref{fig:probability} (top panels).
Below each histogram, we plot the corresponding pairs of coordinates $(x_m,y_{m'})$ as 2D histograms.
%%%%
% Fig. 3 (top panel): collect the configurations from Monte carlo {x_1,x_2,y_1,y_2},
% then keep only the configurations where x_1,x_2 < y_1,y_2 (so we restrict dataset to a conditional probability distribution),
% then we plot all x as the histogram {x_1,x_2,x_1,x_2, ...}
% the same procedure fot y
% overalpping of two densities for x and y, comes from two different configurations, where if we compare e.g. x_2 from fist configuration with y_1 form second then x_2>y_1, but this is fine
%%%%%
Histograms illustrate the probability density of the many-body wavefunction~\eqref{eq:Jastrow}
obtained using optimized coefficients $\{\vec{c}_A,\vec{c}_B \} = \{c_1^A, c_2^A; c_1^B,c_2^B,c_3^B\}$ which minimize the quasi-energy functional~\eqref{eq:quasienergy-functional} for a given intra-species interaction strength $g_A$.

In the symmetry-preserving regime, $g_A = 0$,
the coefficients $\{|\vec{c}_A|^2,|\vec{c}_B|^2\}$ are uniform within each resonance.
The corresponding histograms, shown in Fig.~\ref{fig:probability}(a)-(b),
exhibit interference patterns for both clouds $A$ and $B$ due to atoms occupying multiple Wannier states.
In the symmetry-broken regime at $|g_A| \gtrsim 415$, the optimized coefficients become non-uniform.
We identify two distinct representative cases.
The first case, for $|g_A| = 610$ with optimized coefficients $\{|\vec{c}_A|^2,|\vec{c}_B|^2\} = \{0.46, 0.54; 1, 0, 0\}$. This indicates that atoms in cloud $A$ occupy two Wannier states with slightly different weights, while those in cloud $B$ occupy a single Wannier state. Due to this difference in occupancy, the probability density for cloud $A$ (yellow bars in Fig.~\ref{fig:probability}(c)) exhibits an interference pattern, whereas the probability density for cloud $B$ (gray bars) does not. 
The time evolution of cloud $A$ barely indicates spontaneous breaking of the discrete time translation symmetry, i.e., $|c_1^A|^2\approx |c_2^A|^2$. It means that we observe a situation where time translation symmetry is not broken in the driving force of cloud $B$ (i.e. in cloud $A$)
while time translation symmetry is broken in the driven cloud $B$ itself.
For $|g_A| = 680$, atoms in each cloud occupy only a single Wannier state, with the many-body state~\eqref{eq:Jastrow} 
characterized by $\{|\vec{c}_A|^2,|\vec{c}_B|^2\} = \{1,0;1,0,0\}$. 
This leads to the most pronounced discrete time translation symmetry breaking. 
The probability densities are presented for two time snapshots in Fig.~\ref{fig:probability}: 
one where particles of different types are close in configuration space (Panels (e)-(f)) 
and another where they are farther apart (Panels (g)-(h)).

%%%%%%%%%%%%%%%%%%%%%%%%%%%%%%%%%%%%%%%
%%%%%%%%%%%%%%%%%%%%%%%%%%%%%%%%%%%%%%%
%%%%%%%%%%%%%%%%%%%%%%%%%%%%%%%%%%%%%%%

\section{\label{sec:summary}Summary}
We have analyzed a system of two distinct ultracold atomic clouds with weak attractive intra-species interactions and infinitely strong inter-species interactions. 
The system is placed in a one-dimensional space in the presence of the gravitational field, where the lower cloud bounces on an oscillating mirror, effectively acting as a periodic driving force for the upper cloud due to the infinite inter-species repulsion. 
Our focus was on understanding how intra-species interactions can lead to discrete time translation symmetry breaking, resulting in the formation of a complex DTC.

To study this phenomenon, we employed a Jastrow-like ansatz as a variational many-body wavefunction to identify the low-energy states within the resonant Hilbert subspace. 
Our analysis revealed two distinct regimes: a symmetry-preserving phase, 
where the variational coefficients remain uniformly distributed, 
and a symmetry-broken phase, where these coefficients become non-uniform, signaling the spontaneous breaking of discrete time translation symmetry. Within the latter regime, we observe two sub-regimes: in one, the lower atomic cloud, which provides the driving force for the upper cloud, barely breaks time translation symmetry, while in the other, both clouds exhibit evident symmetry breaking.
The emergence of a complex DTC is characterized by an overall period distinct from the external driving period.

Due to the exponential increase in computational complexity with particle number, 
we restricted our study to a small system with $N_A=N_B=2$ particles. 
However, we expect that similar results --
the emergence of complex DTC -- will hold for larger number of particles. 
Our findings show that, since the bottom cloud serves as an effective periodic driving force for the upper cloud, 
this leads to a cascade of spontaneous symmetry breaking giving rise to a collective discrete time-crystalline phase.

We anticipate that in a stacked 1D configuration with multiple ultracold atomic clouds, where each cloud is effectively driven by the one below it, and the bottommost cloud is driven by an oscillating mirror, a similar cascading effect will occur. Spontaneous discrete time translation symmetry breaking is expected to propagate through the entire system, leading to a cascade of DTCs. 

%%%%%%%%%%%%%%%%%%%%%%%%%%%%%%%%%%%%%%%
%%%%%%%%%%%%%%%%%%%%%%%%%%%%%%%%%%%%%%%
%%%%%%%%%%%%%%%%%%%%%%%%%%%%%%%%%%%%%%%

\section{\label{sec:acknowledgments}Acknowledgments}
We thank Krzysztof Giergiel, Arkadiusz Kuro\'s and Czcibor Ciostek for fruitful discussions. This research was funded by the National Science
Centre, Poland, Project No.~2021/42/A/ST2/00017.

%%%%%%%%%%%%%%%%%%%%%%%%%%%%%%%%%%%%%%%
%%%%%%%%%%%%%%%%%%%%%%%%%%%%%%%%%%%%%%%
%%%%%%%%%%%%%%%%%%%%%%%%%%%%%%%%%%%%%%%

\bibliography{ref.bib}

\end{document}